\documentclass[aps,prl,reprint,groupedaddress]{revtex4-1}

\usepackage{siunitx}
\usepackage{amsmath}
\usepackage{amssymb}
\usepackage{graphicx}
\usepackage{times}
\usepackage{rotating}
\usepackage{bm}
\usepackage{color}

\newcounter{lastnote}
\addtocounter{lastnote}{+1}%

\newcommand{\fcs}{Fe$_{1-x}$Co$_{x}$Si}

\begin{document}

\title{Entropy-limited topological protection of skyrmions}

\author{J. Wild$^1$, T.N.G. Meier$^1$, S. Poellath$^1$, M. Kronseder$^1$,  A. Bauer$^2$, 
		A. Chacon$^2$, M. Halder$^2$,  M. Schowalter$^3$, A. Rosenauer$^3$, J. Zweck$^1$,  
		J. M\"uller$^4$, A. Rosch$^4$, C. Pfleiderer$^2$, C. H. Back$^{1,\ast}$}

\affiliation{$^{1}$Institut f\"ur Experimentelle Physik, Universit\"at Regensburg, D-93040 Regensburg, Germany\\
	$^{2}$Physik-Department, Technische Universit\"at M\"unchen, D-85748 Garching, Germany\\
	$^{3}$Institut f\"ur Festk\"orperphysik, Universita\"at Bremen, Otto-Hahn-Allee 1, D-28359 Bremen, Germany\\
	$^{4}$Institut f\"ur Theoretische Physik, Universit\"at zu K\"oln, D-50937 K\"oln, Germany\\
	$^\ast$To whom correspondence should be addressed; E-mail:  christian.back@ur.de}

\date{\today}

\begin{abstract}\textbf{
	Magnetic skyrmions are topologically protected whirls that decay through singular magnetic configurations known as Bloch points. We have used Lorentz transmission electron microscopy to infer the energetics associated with the topological decay of magnetic skyrmions far from equilibrium in the chiral magnet Fe$_{1-x}$Co$_x$Si. We observed that the life time $\tau$ of the skyrmions depends exponentially on temperature, $\tau \sim \tau_0 \, e^{\Delta E/k_B T}$. The prefactor $\tau_0$ of this Arrhenius law changes by more than 30 orders of magnitude for small changes of magnetic field reflecting a substantial reduction of the life time of skyrmions by entropic effects and thus an extreme case of enthalpy-entropy compensation. Such compensation effects, being well-known across many different scientific disciplines, affect  topological transitions and thus topological protection on an unprecedented level. }
\end{abstract}

\maketitle


A question studied in many fields of the natural sciences concerns the lifetime of metastable states. Thermal activation across energy barriers governs e.g. chemical reactions,  the lifetime of memory elements in computers and in hard disks, and transport of ions and electrons in disordered media. Often such processes are controlled by a characteristic energy, the activation energy. It has, however, also been established that a large number of different pathways across an activation barrier leads to a large entropic correction, reducing the lifetime of metastable states and thus the importance of the energy barriers. This effect is known as enthalpy-entropy compensation in the context of chemistry or Meyer-Neldel rule in material sciences. Enthalpy-entropy compensation has, for example, been observed for catalytic reactions \cite{1925:compensation}, transport in semiconductors \cite{1937:Meyer,2010:semicomp}, biological processes \cite{2001:compensationBio}, and in many other fields \cite{1982:Peacock:PRB,1992:Yelon:PRB}.

In recent years differences of the topology of physical states has been widely portrayed as providing exceptionally high stability. Topology represents a branch of mathematics concerned with those properties of geometric configurations which are unaffected by smooth deformations. Famous examples for topologically non-trivial objects include superconducting vortices, certain magnetic textures, structural defects, and surface states of topological materials. However, despite this abundance of topologically non-trivial configurations in nature, an unresolved key question concerns their stability when being part of an ensemble far from equilibrium. 

Particularly suitable to clarify this issue are skyrmions in spin systems with chiral interactions, because a well-founded highly advanced theoretical understanding exists in excellent agreement with experiment. Representing topologically non-trivial spin-whirls, skyrmions were experimentally first identified in the B20 compounds MnSi and {\fcs} \cite{2009:Muhlbauer:Science,2010:Munzer:PhysRevB}, followed more recently by a wide range of bulk compounds \cite{2012:Seki:Science,2013:Nagaosa:NatureNano,2015:Tokunaga:NatCommun}, surface- and interface-based systems \cite{2011:Heinze:NaturePhys,2016:Boulle:NatNano,2016:MoreauLuchaire:NatNano,2013:Nagaosa:NatureNano}, as well as hetero- and nanostructures \cite{2013:Fert:NatNano,2015:Jiang:Science}. With typical dimensions  from a few up to several hundred nanometers, skyrmions in magnetic materials are accessible to a wide range of experimental techniques. Moreover, they are also of immediate interest for spintronics applications. This concerns at present foremost memory elements \cite{2013:Fert:NatNano,2017:twolaneJan,2017:Hsu}, where lifetimes exceeding 10 years represent the technical requirement. In turn, design of metastable states with long lifetimes are both mandatory and a major motivation for the study reported here.

\begin{figure}[t]
	\begin{center}
		\includegraphics[width=\linewidth]{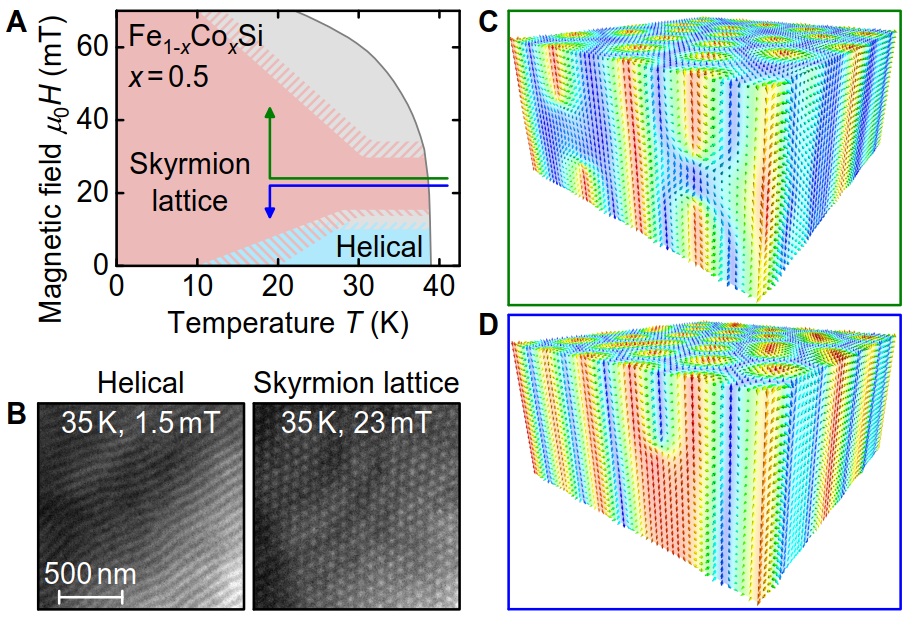}
	\end{center}
	\caption{ (A) Magnetic phase diagram of  {\fcs} with $x=0.5$ obtained by first cooling the system at fixed magnetic field, $B\approx  23$\,mT, and the raising or lowering the field at fixed temperature $T$.  The decrease of the applied field triggers the decay into a helical configuration while either a conical or ferromagnetic state is reached for increasing field. (B) Typical Lorentz-force transmission electron micrscopy images of helical and skyrmion lattice order, respectively. (C) Schematic picture of an early state of the decay of the skyrmion lattice towards a  ferromagnetic state. The skyrmion splits by the formation of a pair of Bloch points located at the end of the skyrmion strings which move towards the surface. (D) Decay of skyrmion lattice order towards the helical state. Neighboring skyrmions merge and a Bloch point
		at the merging points moves towards the surface.
		\label{fig1}}
\end{figure}


For our study we have chosen the B20 compound {\fcs} with $x=0.5$, which displays a  well-understood bulk phase diagram that is generic for this class of materials \cite{2009:Muhlbauer:Science,2010:Munzer:PhysRevB,2013:Buhrandt:PRB,2016:Bauer:Book}. With decreasing temperature, a paramagnetic to helimagnetic transition occurs at $T_{\rm c}$.  
A lattice of skyrmion lines forms in a small field and temperature window just below $T_{\rm c}$ as first established by small-angle neutron scattering \cite{2010:Munzer:PhysRevB}. Lorentz force transmission electron microscopy (LTEM) in {\fcs} provided real space images of skyrmions \cite{2010:Yu:Nature}. While the skyrmion phase in bulk materials is thermodynamically stable in a small pocket of phase space only, it was shown early on that  the range of stability, albeit being hysteretic, increases dramatically and extends down to zero temperature in ultra-thin bulk samples 
 \cite{2010:Yu:Nature,2013:Rybakov,2015:Kiselev:PRL}.
A  typical phase diagram obtained in our sample when cooling at a fixed magnetic field and followed by an increase or decrease of the field at fixed low temperature is shown in Fig. \ref{fig1}\,(A). In the red shaded region the skyrmion lattice is stable on experimentally relevant time scales, but decays when the magnetic field is increased or decreased further.

Schematic depictions of typical spin configurations during early stages of the decay into a ferromagnetic state under increasing field or into a helical state under decreasing field are shown in Fig. \ref{fig1}\,(C) and (D), respectively. The decay into the helical state was first addressed by means of magnetic force microscopy (MFM) of the surface of a bulk sample of {\fcs} ($x=0.5$) \cite{2013:Milde:Science}. Here the decay was found to occur by a merging of skyrmions, implying the presence of a Bloch point that acts like a zipper between skyrmions, i.e., a point of vanishing magnetization enabling the unwinding of the non-trivial topology (Fig.\,\ref{fig1}\,(D)). In comparison, the decay into the conical or ferromagnetic state is theoretically predicted to take place by a pinching-off through the creation of a pair of Bloch-points as illustrated in Fig.\,\ref{fig1}\,(C), and the subsequent motion of the Bloch points towards the surface \cite{2014:Schuette:PRB,2015:Kiselev:PRL}. 

Several studies had reported hysteretic and metastable skyrmion states. Early SANS studies on bulk samples of {\fcs}\,\, revealed the possibility to super-cool the skyrmion lattice into a metastable state \cite{2010:Munzer:PhysRevB,2016:Bauer:PRB}. Detailed magnetization and ac susceptibility measurements as combined with SANS showed \cite{2017:Bauer:preprint}, that the super-cooled skyrmion lattice order for temperatures below $\sim 10\,{\rm K}$ may only be destroyed by means of an applied magnetic field
exceeding the conical to ferromagnetic transition field, i.e., at low temperatures thermal activation is not sufficient to trigger a decay of the metastable skyrmion state. 
The observation of sizeable super-cooling effects in MnSi under pressure \cite{2013:Ritz:PhysRevB} or after violent quenches exceeding 400\,K/min \cite{2016:Oike:NaturePhys}, and, in Cu$_{2}$OSeO$_{3}$, under electric field cooling \cite{2016:Okamura:NatComm}, and the observation of slow relaxation at low temperatures in GaV$_4$S$_8$  \cite{2017:Kezsmarki}, underscore the generic existence of super-cooling effects noticed first in {\fcs}. Several studies also investigated skyrmion generation and destruction triggered by electric currents \cite{2013:Romming:Science,2016:Woo:NatMater,2015:Jiang:Science}. The energy barrier for skyrmion creation/annihilation in two dimensions was studied theoretically in~\cite{2015:Bessarab,Hagemeister2015}, complementing studies in three dimensions \cite{2014:Schuette:PRB,2015:Kiselev:PRL}. An elegant approach to explore the energetics requires time resolved real space imaging of large ensembles.

In our experimental studies we used time-resolved Lorentz force transmission electron microscopy (LTEM) on thin bulk samples of {\fcs}. On the one hand this method provides the required spatial and temporal information, without driving the skyrmion decay.  On the other hand and as explained above, {\fcs} represents an extremely well-understood and well-characterized material suitable to address these issues.

For our measurements a sample with a thickness of $d\approx240\,{\rm nm}$ was prepared from the same optically float-zoned single crystal studied before using a focussed ion beam\cite{2013:Milde:Science,2016:Bauer:PRB,2017:Bauer:preprint}. The single crystal was cut such that a $\langle 100\rangle$ axis was normal to the platelet. The magnetic field was applied along the same direction. As shown in Fig.\,\ref{fig1}\,(B), for zero magnetic field and low temperatures well-defined helical order is observed with a modulation length $\lambda\approx90\,{\rm nm}$, in agreement with previous reports. Further, under field-cooling in an applied field of 23\,mT a well defined hexagonal skyrmion lattice forms just below $T_{\rm c}$. When reducing the temperature further, the same unchanged skyrmion lattice persists down to the lowest temperatures studied ($ 12\,{\rm K}$).

\begin{figure}[t]
	\begin{center}
		\includegraphics[width=\linewidth]{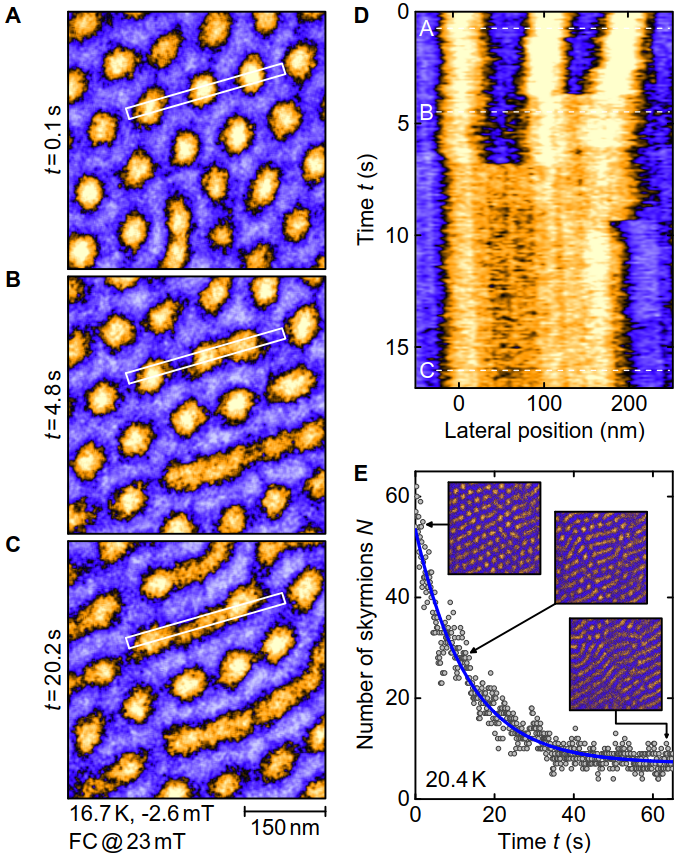}
	\end{center}
	\caption{Key characteristics of the decay of skyrmions into helical order. 
		The sample was field-cooled (FC) from above the helical transition temperature ($T_c\approx38\,{\rm K}$) under an applied magnetic field $B=23\,{\rm mT}$ down to $T_{\rm m}$, where the field was reduced to $B_{\rm m}$ and data recorded as a function of time $t$. 
		(A), (B), and (C) Typical LTEM patterns at $T_{\rm m}=16.7\,{\rm K}$ after reaching $B_{\rm m}=-2.6\,{\rm mT}$ for $t=0.1\,{\rm s}$, 4.8\,s, and 20.2\,s, respectively. (D) Evolution of the intensity across the white box marked in panels (A), (B), and (C) as a function of time (vertical axis). 
		(E) Typical time dependence of the number of skyrmions for $T_{\rm m}=20.4\,{\rm K}$ and $B_{\rm m}=-2.6\,{\rm mT}$. The blue curve represents an exponential fit.
		\label{fig2}}
\end{figure}

In our LTEM measurements the evolution of the magnetic state was studied with a time resolution of $\sim100\,{\rm ms}$, where movies were recorded after field-cooling and a subsequent field change. We first investigate the destruction of the skyrmion state when the magnetic field is decreased. Shown in Fig.\,\ref{fig2} are typical data for a decay into the helical state, observed after field-cooling at 23\,mT down to $T_{\rm m}=16.7\,{\rm K}$ and a reduction of the field to $B_{\rm m}=-2.6\,{\rm mT}$. As illustrated in Figs.\,\ref{fig2}\,(A), (B) and (C) for patterns recorded at $t=0.1\,{\rm s}$, $4.8\,{\rm s}$ and $20.2\,{\rm s}$ after reaching $B_{\rm m}$, respectively, the intensity pattern displays a merging of skyrmions. 
This  process corresponds accurately to the mechanism observed by means of MFM on the surface of bulk specimens cut from the same single crystal \cite{2013:Milde:Science}. 

The merging of skyrmions is additionally illustrated in Fig.\,\ref{fig2}\,(D), which displays the intensity across the white box in panels (A), (B), and (C) as a function of time along the vertical axis. Details of the merging cannot be resolved for the frame rate of our experiments, i.e., the motion of the monopoles across the thickness of the sample of 250\,nm is faster than 100\,ms corresponding to a speed greater $2.5\cdot \,10^{-6}\,{\rm m/s}$. However, using a bespoke algorithm we could track the number of skyrmions $N$ for a given area as a function of time as illustrated in Fig.\,\ref{fig2}\,(E). As a main new result, we observe an exponential time dependence, 
$N \approx N_{0} e^{-t/\tau(B,T)}$ (blue line), from which we  extract the lifetime $\tau(B,T)$ as function of field and temperature, analyzed further below. Note that $N_{0}$ is in general smaller than the initial number of skyrmions, as some skyrmions have already decayed during the field sweep when only blurred images are recorded due to image drifts.

\begin{figure}
	\begin{center}
		\includegraphics[width=0.94\linewidth]{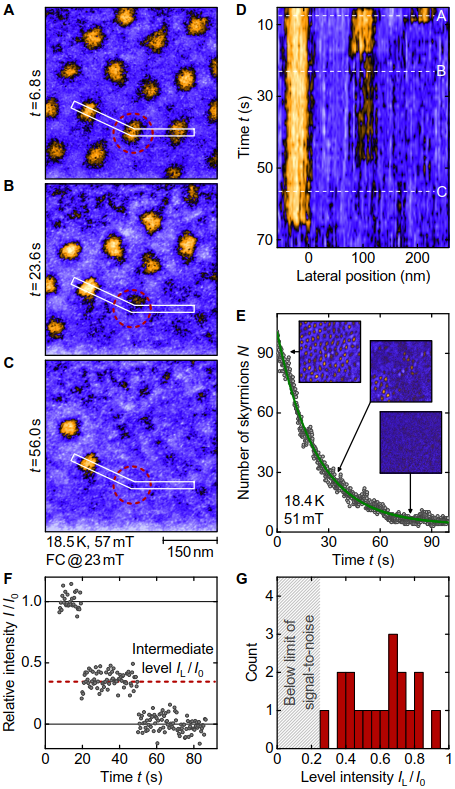}
	\end{center}
	\caption{
		Key characteristics of the decay of skyrmions for increasing magnetic fields.
		The sample was field-cooled (FC) from above the helical transition temperature ($T_c\approx38\,{\rm K}$) under an applied magnetic field $B=23\,{\rm mT}$ down to $T_{\rm m}$, where the field was increased to $B_{\rm m}$ and data recorded as a function of time $t$. 
		(A), (B), and (C) Typical LTEM patterns at $T_{\rm m}=18.5\,{\rm K}$ after reaching $B_{\rm m}=57\,{\rm mT}$ for $t=6.8\,{\rm s}$, 23.6\,s, and 56.0\,s, respectively. (D) Evolution of the intensity across the white box marked in panels (A), (B), and (C) as a function time (vertical axis). 
		(E) Typical time-dependence of the number of skyrmions for $T_{\rm m}=20.4\,{\rm K}$ and $B_{\rm m}=57\,{\rm mT}$, fitted by an exponential (green line).
		(F) Time dependence of the intensity within the red-dashed circle in panels (A), (B), and (C). For a small number of skyrmions, a two-step decay via an intermediate state with lower intensity is observed. (G) Statistics of the intermediate-state intensities. 
		\label{fig3}}
\end{figure}

The destruction of the skyrmion state after field-cooling, when a magnetic field increase triggers a decay into a conical (or ferromagnetic) state, are summarized in Fig.\,\ref{fig3}. Data shown were recorded after field-cooling at 23\,mT down to $T_{\rm m}=18.5\,{\rm K}$ and an increase of the field to $B_{\rm m}=57\,{\rm mT}$. As predicted by theory  \cite{2015:Kiselev:PRL}, the decay pattern is characterized by the disappearance of individual skyrmions, as opposed to the merging observed for a decay into the helical state. 
 Fig.\,\ref{fig3}\,(D) displays the intensity across the white box in panels (A), (B) and (C) as a function of time along the vertical axis. Using the same algorithm to track the number of skyrmions again an exponential time dependence is observed for the entire parameter range accessible, as illustrated in Fig.\,\ref{fig3}\,(E). Both the qualitative decay mechanism, as well as the specific time dependence analyzed below represent main results of our study. 

We note that for most of the decays individual skyrmions vanish suddenly between two frames of our movies. However, for less than $10\%$ of the skyrmions the intensity does not vanish in a single step but exhibits a two-step process. An example is marked by the red-dashed circle in Figs.\,\ref{fig3}\,(A), (B), and (C), corresponding to a diameter of $\sim 100\,{\rm nm}$ in Fig.\,\ref{fig3}\,(D). 
The relative change of intensity in this region, $I/I_0$, as a function of time is shown in Fig.\,\ref{fig3}\,(F). The observation of the intermediate level implies, that part of the skyrmion survives as an intermediate, metastable skyrmion string with a length shorter than the thickness of the sample. At least one end of the skyrmion string is therefore inside the sample and topology enforces the presence of a Bloch point at this location. In turn, the metastable state consists of at least one Bloch point and a skyrmion string which connects either one Bloch point (or two) to the surface or connects a pair of Bloch points which each other.

One of the most likely mechanisms causing the metastable intermediate state is trapping of the Bloch points by local defects. As an interesting alternative, the metastable state may be a so-called chiral bobber, predicted theoretically by Rybakov {\it et al.} \cite{2015:Kiselev:PRL}. The authors of this study pointed out that the surface energy of the skyrmion provides a repulsive potential for the Bloch point. Thus a chiral bobber represents a Bloch point located immediately below the surface of the sample. To clarify the nature of the metastable states we observed, we analyzed the intensity of 18 of these inhibited decays (out of the 355 decays we investigated in detail). Plotting the number of inhibited decays as a function of relative intermediate intensity, shown in Fig.\,\ref{fig3}\,(G), we observe a very broad distribution. Taking these intensities as a measure of the length of the intermediate states, most of them are larger than expected theoretically for a single bobber located close to the surface. Hence, within our limited statistics, an interpretation in terms of Bloch points trapped by defects appears to be the most likely scenario.

\begin{figure}
	\begin{center}
		\includegraphics[width=\linewidth]{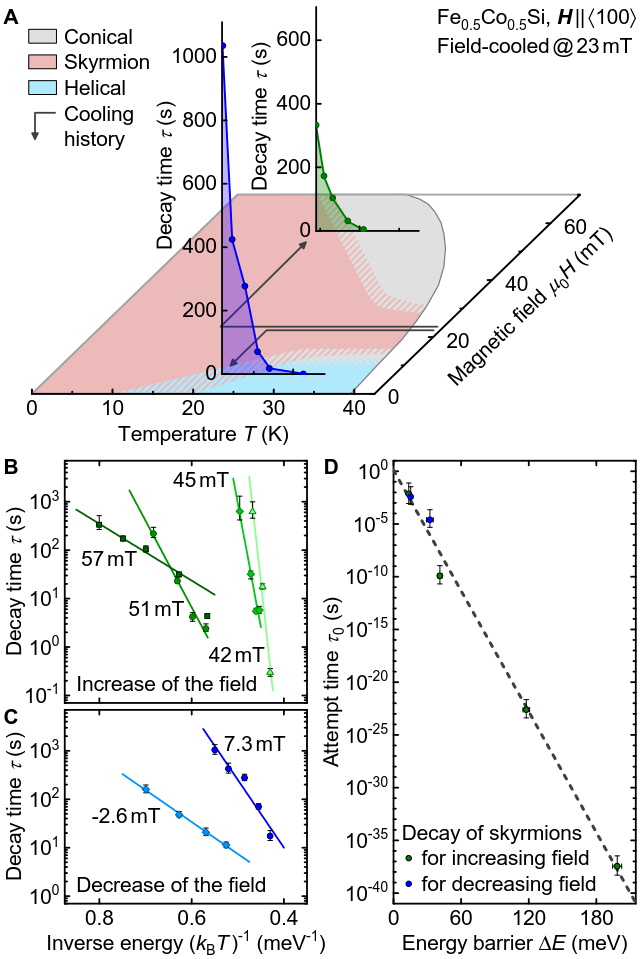}
	\end{center}
	\caption{Key characteristics of the decay rates of supercooled skyrmions in {\fcs} ($x=0.5$). 
		(A) Typical decay times $\tau$ after field-cooling at $B=23\,{\rm mT}$, followed by a decrease/increase to $B_{\rm m}$. (B) and (C) Decay time $\tau$ as a function of thermal energy for increasing and decreasing magnetic field, respectivly, and various values of $B_{\rm m}$.
		(D) Attempt time $\tau_0$ as a function of energy barrier $\Delta E$ inferred from the exponential time dependence of the skyrmion decay.  The variation of more than 30 orders of magnitude of $\tau_0$ reflects extreme enthalpy-entropy compensation. 
		\label{fig4}}
\end{figure}

Taking together all of our data, the lifetime $\tau(B,T)$ of skyrmions depends sensitively on magnetic field and temperature. Shown in Fig.\,\ref{fig4}\,(A) are typical decay times as a function of temperature for selected field values. The phase diagram shown in the background of that figure is identical to  Fig.\,\ref{fig1}\,{A}. On a double-logarithmic scale the decay times as a function of thermal energy display strong variations with magnetic field, where the decay into the conical and helical state are shown in Figs.\,\ref{fig4}\,(B) and (C), respectively. 
As expected, the temperature dependence can  be described approximately by an Arrhenius law, $\tau(B,T) \approx \tau_0(B) \, e^{\Delta E(B)/k_B T}$. However, the energy barriers vary strongly with the strength of the magnetic field, being largest close to the stability regime of the skyrmion lattice. For increasing field the energy barrier therefore drops, e.g., from $\Delta E = (199 \pm 4)$~meV for $B_{m} = 42.2$~mT down to $\Delta E = (13 \pm 1)$~meV for $B_{m} = 56.8$~mT. When the field is reduced and the skyrmions decay into the helical state, the activation energy $\Delta E = (32 \pm 3)$~meV at $B_{m} = 7.3$~mT close to the stability region is again larger than further away, where we find $(15 \pm  1)$~meV for $B_{m} = -2,6$~mT.

This qualitative trend is consistent with existing theoretical predictions. Namely, in Ref.~\cite{2014:Schuette:PRB} stochastic micromagnetic simulations at finite temperatures were used to investigate the activation energy and magnetic field dependence for the decay of a single skyrmion into the helical state. Similar to our experiments a strong dependence on the magnetic field $B$ was found: notably the decay rate increases rapidly when the field is lowered. Further, Rybakov {\it et al.}  \cite{2015:Kiselev:PRL} used a variational approach at $T=0$ to estimate activation barriers for a transition into a conical state. As in our experiments, theoretically predicted activation energies for this transition exceed those for the transition into the helical state. Using straightforward scaling arguments, see supplementary material,  to extrapolate the calculated activation energies to the experimentally relevant parameter regime, we find, however, that the measured activation energies are about an order of magnitude smaller than the predicted ones.
 At present, we have no explanation for this substantial discrepancy, however, we speculate that it is connected with the large entropic effects discussed below which become most pronounced for large activation barriers.
 
Surprisingly, measurements of the activation energies alone do {\em not} allow to predict the lifetime of the metastable skyrmion state. Instead we find that the prefactor $\tau_0(B)$ -- the so-called attempt time -- of the Arrhenius dependence assumes extremely small values and shows an unusually strong sensitivity to magnetic fields. This is emphasized in Fig.\,\ref{fig4}\,(D), which shows that $\tau_0$ as function of the activation energy changes from extremely low values, smaller than $10^{-37}$\,s,  for the measurements at $B_m=42.2 \; \mathrm{mT}$, where the largest activation energy was measured, to almost macroscopic time scales $\sim 10^{-2}$\,s at $B_m=57 \; \mathrm{mT}$. These values should be contrasted with typical microscopic time scales $\sim10^{-9}\,{\rm s}$ commonly accepted for estimates of attempt times in magnetic materials~\cite{1999:Weller:IEEE,chen2010,1994:Lederman:PRL}.
 
To account for this extreme variation, we revisit the Arrhenius-law used for fitting the decay times. At finite temperature, thermodynamics and transition rates are governed by the free energy $F(T,B)=E(T,B)-T\cdot S(T,B)$. Thus, the Arrhenius-type decay law for a single energy barrier assumes the form
$\tau(B)=\tau_{00} \exp\left( \frac{\Delta F(T,B)}{k_\mathrm{B} T} \right)$, where $\Delta F=F_b-F_0$ is the free-energy difference of the initial state, $F_0$, and a highly excited state, $F_{b}$, which defines the bottleneck of the skyrmion decay.
Inserting the definition of the free energy one obtains
\begin{equation}
\tau(B)=\tau_{00} \exp\left(-\frac{\Delta S(B)}{k_B}\right) \exp\left( \frac{\Delta E(B)}{k_\mathrm{B} T} \right)
\end{equation}
In turn, the high-energy offset of the decay time, given by $\tau_0=\tau_{00} \exp(- \Delta S/k_B)$, depends exponentially on the entropy difference $\Delta S$ between the state with skyrmion and the bottleneck state. A large positive value of $\Delta S$ leads to an exponential reduction of $\tau_0$ and strongly increases the transition rates. Physically, this is due to the exponentially large number $N_p$ of microscopic pathways across the energy barrier, $\Delta S =k_B  \ln[N_p]$, which increases the transition rate by the factor $N_p$. Similarly, a negative value of $\Delta S$ takes into account a reduction of the transition rate, which arises when the number of microscopic realizations of the initial state with skyrmion is much larger than the number of states close to the bottleneck.

Entropic effects as discussed here occur in many different systems in biology, chemistry, and physics. As the high-energy states typically have a larger entropy, these entropic effects tend to counterbalance partially the energetic effects. Therefore this phenomenon is often referred to as 'compensation effect'. Phenomenologically, linear relationships of activation energy and entropy are known as Meyer-Neldel rule in the context of solid state physics\cite{1937:Meyer,1982:Peacock:PRB,1992:Yelon:PRB}. In the limit where the activation energy is much larger than the typical microscopic energy scale $E_m$, they arise \cite{1982:Peacock:PRB,1992:Yelon:PRB} because the number of ways the activation barrier can be reached by microscopic processes varies exponentially in $\Delta E/E_m$, $N_p \sim e^{\Delta E/E_m}$ and hence $\Delta S\sim k_B \frac{\Delta E}{E_m}$. The fit in Fig.\,\ref{fig4}\,(D)  shows that $\tau_0$ follows the Meyer-Neldel rule approximately and exhibits an exponential dependence on the activation energy $\Delta E$. From the fit we obtain $E_m/k_B \approx 27\pm 3$\,K which is indeed of the correct order of magnitude expected for our system where $T_c \approx 38\,{\rm K}$. Remarkably, the approximate linear relation of $\Delta S$ and $\Delta E$ extends even in the regime where $\tau_0$ is much larger than microscopic time scales, strongly suggesting a negative $\Delta S$.



Our study reveals unprecedentedly large compensation effects by more than 30 orders of magnitude for the decay of topologically non-trivial spin configurations, notably skyrmions, as part of an ensemble far from equilibrium. This may seem particularly relevant in view of the technological impact expected of topological protection. It concerns, for instance, specifically the long-term stability of magnetic memory devices inferred from measurements of activation energies. Whereas typical values of $\tau_0 \sim10^{-9}\,{\rm s}$ are broadly accepted \cite{1999:Weller:IEEE,chen2010}, our results show that, especially for the large activation energies required for long-term stability, such estimates may overestimate the thermal stability of skyrmion-based memory devices by factors of the order of $10^{20}$ and more. This illustrates, that entropic effects will be very important for the technological exploitation of topologically non-trivial systems such as memory technology based on skyrmions 
\cite{acknowledgements}.

\newpage

\bibliographystyle{Science}



\clearpage

\onecolumngrid
\appendix

\begin{center}\Large\textbf{{Supplementary Materials for \\ Entropy-limited topological protection of skyrmions}}\end{center}

\section{Sample preparation}

\begin{figure}
	\centering
	\includegraphics[width=0.7\linewidth]{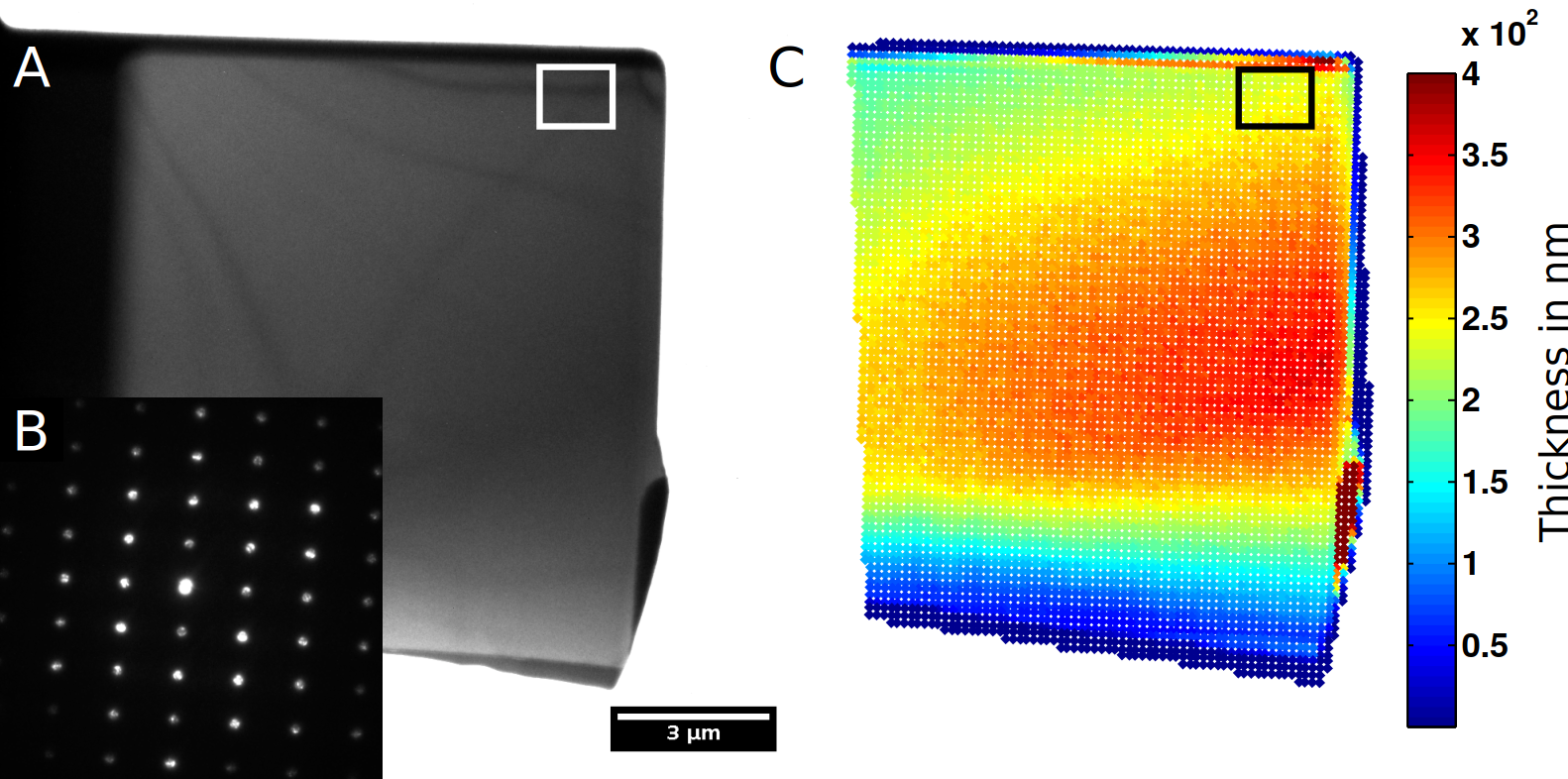}
	\caption{(A) shows the FIB prepared sample with a dimension of $\SI{10}{} \times \SI{10}{\micro\meter}$. A diffraction pattern in (B) confirms the thinning of the specimen in [100] direction. The thickness extracted from HAADF-STEM in (C) gives a thickness of \SI{241 \pm 8}{nm} for the measurement position.}
	\label{fig:Figure_thickness}
\end{figure}

A large single crystal of Fe$_{1-x}$Co$_{x}$Si with $x = 0.5$ was grown by means of the optical floating-zone technique under ultra-high vacuum conditions~\cite{2011:Neubauer:RevSciInstrum, 2016:Bauer:RevSciInstrum2, 2016:Bauer:PhysRevB}. From the single-crystal ingot, we cut a bulk sample of a thickness of \SI{200}{\micro\meter} along a $\langle100\rangle$ axes using X-ray Laue diffraction and a wire saw. This platelet was thinned by the focused ion beam (FIB) milling technique in plane-view configuration and subsequently polished by Ar-ions at low voltages. The finished TEM lamella is shown in Fig. \ref{fig:Figure_thickness} (A) with dimensions of $\SI{10}{}\times\SI{10}{\micro\meter}$. A TEM diffraction pattern in Fig.  \ref{fig:Figure_thickness} (B) confirms the thinning of the specimen in $\langle100\rangle$ direction. 
The thickness of the specimen was determined by high angle annular dark field scanning TEM (HAADF-STEM) imaging. The measured signal was normalized to the incoming beam intensity extracted from a detector scan \cite{Rosenauer}. The normalized intensity was then compared with frozen phonon multislice simulations yielding the local specimen thickness. A thickness map is shown in Fig. \ref{fig:Figure_thickness} (C). All measurements were performed in the marked area with a thickness of \SI{241 \pm 8}{nm}.

It is interesting to note that the thickness of our sample exceeds the modulation length of the helical state by a factor of 2.7. In comparison to bulk samples cut from the same ingot, which show a helimagnetic transition temperature of $T_{\rm c}=47\,{\rm K}$, the ordering temperature observed in the thin sample was $T_{\rm c}=38\,{\rm K}$. This reduction corresponds very well with a similar effect observed in MnSi, which was attributed to an enhancement of fluctuations with reduction of sample thickness
Moreover, in the study on MnSi a skyrmion lattice was observed across the entire magnetic phase diagram~\cite{2012:Tonomura:NanoLett}, consistent with LTEM studies in FeGe, where an increase of the temperature and field range of the skyrmion lattice was observed with a reduction of sample thickness~\cite{2010:Yu:Nature,2011:Yu:NatureMater}. 

\section{Magnetic imaging}

Magnetic images were taken with a  FEI Tecnai F30 transmission electron microscope in Lorentz mode (LTEM) where the magnetic field normal to the sample surface is tuned by the objective lens current. A defocused image is projected onto a phosphorescence screen which is filmed by a high-speed camera through a lead glass window. We cooled and controlled the temperature of the specimen with a Gatan liquid Helium holder. 

\section{Image processing for the skyrmion decay measurements}

Within this section the analysis of the LTEM-images for the skyrmion decay measurements is described. The image processing approach can be divided into the preprocessing and the extraction of physical quantities from the preprocessed images. For the two cases of the skyrmion decay to the helical phase and the decay to the conical phase two completely different image processing approaches were developed. Before describing these methods in detail first the preprocessing steps, which both methods have in common are explained.

\subsection{Preprocessing}

The raw LTEM-images not only contain information about the magnetic texture of the sample, but also nonmagnetic information for example due to crystal contrasts. The first step is hence to subtract the nonmagnetic background from the images. Since the nonmagnetic contrast is mostly present on much larger length scales than the helix pitch length or the size of a skyrmion, it can be removed from the images by subtracting a Gaussian filtered version of the image, where the sigma of the Gausian kernel has to be chosen much larger than the skyrmion size. The signal-to-noise ratio of these images can be further enhanced by applying a Fourier-filter using a suitable bandpass-like filter mask to keep only objects with the lateral dimensions of a skyrmion.

\subsection{Analysis of the skyrmion decay to the conical phase}

\begin{figure}
	\centering
	\includegraphics[width=\linewidth]{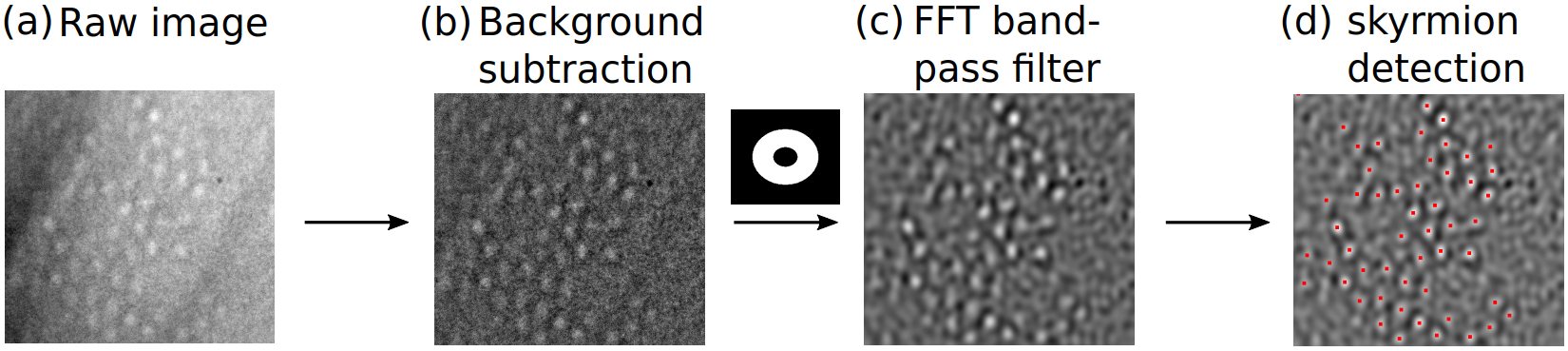}
	\caption{Illustration of the evaluation of the skyrmion decay to the concical phase. (a) shows raw data, (b) the image after background subtraction and (c) a bandpass Fourier-filtered version, which is used for the local maximum detection in step (d) resulting in the skyrmion positions.}
	\label{f_det_conical}
\end{figure}

In the case of the decay to the conical phase the skyrmions are counted in each frame of the video by an automated detection approach, which is illustrated in Fig. \ref{f_det_conical}. The algorithm searches for local maxima of the intensity in each of the images. Each local maximum is considered as a candidate for a skyrmion and is further tested by computing a "local threshold" by comparing the image intensity at the possible skyrmion position with the intensity in a larger area around this position. If the skyrmion candidate exceeds a certain threshold value it will be counted as a skyrmion.

\subsection{Analysis of the decay to the helical phase - skyrmion number method}

The decay of the skyrmions to the helical phase is investigated by two different methods. One of them relies on counting the skyrmion number versus time as the skyrmions merge into the stripes of the helical phase. The other approach is based on tracing the intensity of the two peaks of the helical phase in reciprocal space versus time and is described within the next subsection.

\begin{figure}
	\centering
	\includegraphics[width=\linewidth]{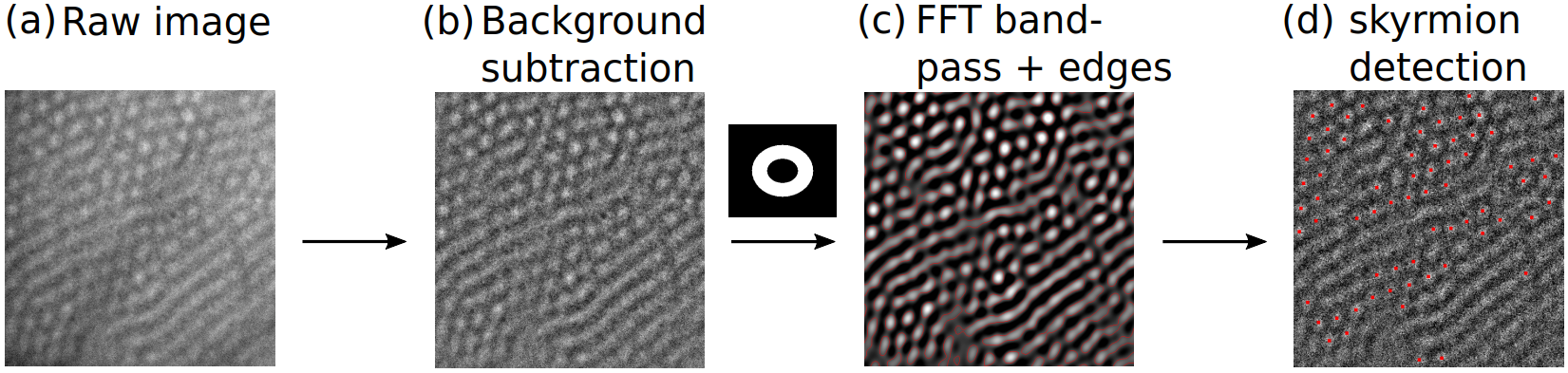}
	\caption{Illustration of the evaluation of the skyrmion decay to the helical phase. (a) shows raw data, (b) the image after background subtraction, (c) a bandpass Fourier-filtered version with contours obtained by the LoG-edge detection algorithm. In (d) the skyrmion positions obtained from the contour image are visualized by red squares.}
	\label{f_det_helical}
\end{figure}

In contrast to the previous subsection the skyrmion counting approach cannot be based on finding local maxima of the intensity, since not only the skyrmions but also the stripes of the helical phase would contribute local maxima. Hence a different skyrmion detection approach was developed and is illustrated in Fig. \ref{f_det_helical}. It relies on finding lines of equal intensity within the video frames by an edge detection algorithm using the Laplacian-of-Gaussian(LoG)-method. After processing with the edge detection skyrmions can be identified as circular contours with a certain enclosed area. The skyrmions are detected by identifying the connected components in the inverted contour image and subjecting each connected component to certain test criteria. The first criterium is that the area enclosed within the contour lies in a certain range, the second criterium is that the eccentricity of the contour has to be low enough (skyrmions are circular) and the final criterium is that the local threshold of the skyrmion compared to its environment is large enough.

\subsection{Analysis of the decay to the helical phase - order method}

The second method for the evaluation of the transition to the helical phase relies on an analysis of the video frames in reciprocal space. In the skyrmion phase the Fourier-transformed image shows six spots. During the transition intensity is transferred from the six spots into the two spots of the helical phase. By defining a ring-shaped mask and comparing the intensity (averaged absolute value of the FFT) of the helical spots with the intensity of the complement of these spots with respect to the ring-shaped mask, while normalizing to the overall intensity, an order parameter can be defined as follows:

\begin{equation}
A_\text{FFT}=\frac{I_\text{angle}-I_\text{compl}}{I_\text{angle}+I_\text{compl}} \ .
\end{equation}

\begin{figure}
	\centering
	\includegraphics{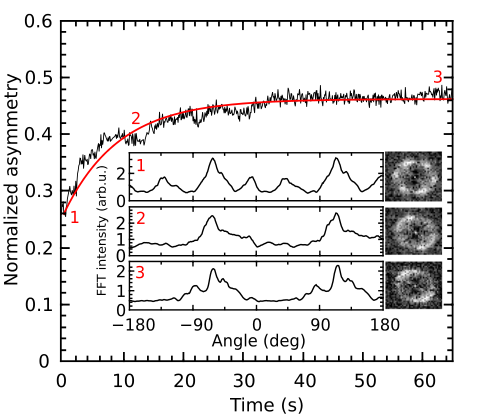}
	\caption{Reduction of skyrmion order deduced from the Fourier transform of the real space data after the magnetic field has been reduced to B$_{ext}$=-2.6~mT from an initial field value of B$_{ext}$=23~mT. The insets show radial averages at distinct times over the FT data shown at the right. The measurement temperature is T$_m$=20.4~K.}
	\label{Figure_FFT_decay}
\end{figure}

This order parameter can be evaluated for each frame in the movies. Since during the transition to the helical phase intensity from the six spots of the skyrmion lattice is transferred into the two spots of the helical phase, the order parameter $A_\text{FFT}$ will increase with time as shown in Fig. \ref{Figure_FFT_decay}. Note that the decay time for skyrmion order is shorter than the decay time for the reduction of the number of skyrmions.

\section{Evaluation of intermediate states}

The evaluation of the intermediate state intensity levels which were observed during the skyrmion decay into the conical phase was done as follows. First, the temporal average intensity of the spatially static skyrmion, intermediate state and conical phase is calculated. Second, a linescan through the maximum intensity of the respective spin object is taken and fitted by a Gaussian function, which gives the skyrmion and intermediate state intensity $I_\text{Sk}$ and $I_\text{i}$. The intensity level of the conical background $I_\text{con}$ is calculated by averaging within an area of the former skyrmion size. The levels shown in Fig. 3 (F) of the main text are then calculated by $I_\text{L}/I_\text{0}=(I_\text{i}-I_\text{con})/(I_\text{Sk}-I_\text{con})$. The temporal linescan in Fig. 3 (G)  of the main text is obtained by averaging the intensity in a small circular area of approximately half the skyrmion diameter around the skyrmion core position that is tracked in time.

\section{Scaling analyis of activation energies}
As the size of skyrmions is much larger than lattice constants, one can efficiently describe their proporties in the continuum limit by a field theory. Such a description will, however, break down in the center of a Bloch point where the spin configuration is singular. 
Within a non-linear sigma model the energy of magnetic textures is computed from
\begin{eqnarray}
E=\int d^3 r \, A (\boldsymbol{\nabla} \hat n)^2+ D\,  \hat n \cdot \boldsymbol{\nabla} \times \hat n - \mu_0 H M \hat n_3 -\frac{\mu_0 M}{2} {\bf H}_{D} \hat n
\end{eqnarray}
where ${\bf H}_{D}(r)=-\frac{M}{4 \pi} \boldsymbol{\nabla} \int  d^3 r' \, 
\frac{r-r'}{|r-r'|^3} \hat n(r')$ is the demagnetization field from dipole-dipole interactions.
Skyrmion formation is governed by the length scale $2A/D$. Therefore, we rewrite the energy by expressing all distances in this length scale. In the resulting dimensionless units, we label coordinates by $\bf x$ with ${\bf r}=\frac{2 A}{D} \bf x$ and the wavelength of the helical state is $2 \pi$. The thickness $d_s$ of the sample can be expressed by the dimensionless  $d=d_s D/(2 A)$. The energy functional then takes the form
\begin{eqnarray}
E=\frac{4 A^2}{D} \int d^3 x \,  \frac{1}{2} (\boldsymbol{\nabla} \hat n)^2+   \hat n \cdot \boldsymbol{\nabla} \times \hat n -h  \hat n_3 - \frac{\tau_D}{2} \hat n \cdot {\bf h_D}
\end{eqnarray}
with the dimensionless demagnetization field ${\bf h}_{D}(x)=-\frac{1}{4 \pi} \boldsymbol{\nabla} \int  d^3 x' \, 
\frac{x-x'}{|x-x'|^3} \hat n(x')$, controlled by two dimensionless parameters $h$ and $\tau_D$, where $h=\frac{ 2 A \mu_0  M}{D^2} H$ is the dimensionless magnetic field measured in units such that the transition from the conical to ferromagnetic state takes place at $h=1$ and 
$\tau_D=\frac{2 A \mu_0  M^2}{D^2}$ is a dimensionless measure for the strength of the dipole-dipole interaction which  typically takes values of order one  (with, e.g.,  $\tau_D=0.65$ for Fe$_{0.8}$Co$_{0.2}$Si \cite{Schwarze}).

The activation energy $\Delta E$ for the decay of the skyrmion is determined by a high-energy state containing a pair of Bloch points \cite{Schuette,Rybakov}. The continuum description breaks down at the core of these singular spin configurations. This regime contributes the core energy $E_c$ which depends on microscopic details. In a metal, for example, the details of the electronic band-structure will enter. Nevertheless, one can estimate that this energy is of the order of the exchange coupling $J$, which can be estimated from the spin-stiffness $A$ using $E_c \sim J \sim A a$, where $a$ is the relevant microscopic length scale (e.g. distance of magnetic ions).

These arguments show that the activation energy can be written in the form
\begin{eqnarray}
\Delta E=\frac{4 A^2}{D} f(h,\tau_D,d) +O(A a) = \frac{J N_s}{2 \pi}  f(h,\tau_D,d)+ O(J)
\end{eqnarray}
where $f$ is a dimensionless scaling function which depends on the dimensionless parameters $h$, $\tau_d$ and $d$ and is independent of the cutoff parameter $a$ in $d=3$. For the second equality, we identified $A=J/(2a)$ with the exchange coupling $J$ of a classical Heisenberg model on a cubic lattice with lattice constant $a$ and introduced $N_s=4 \pi A/(D a)$, the number of spins per pitch of the helix. 
Importantly, for Fe$_{0.5}$Co$_{0.5}$Si, where the pitch of the helix is $\lambda \sim 90$\,nm,  the universal first term arising from the continuum model, is expected to exceed the non-universal second term by a large factor of order $10^2$  for temperatures $T\ll T_c$ where the above analysis applies.

Previous theoretical results computed $\Delta E$ for $N_s \sim 10$ and $\tau_D=0$ \cite{Schuette,Rybakov}. Assuming that the numerical results were dominated by the universal contribution, we can estimate $f(0,0,31)\approx 3.6$ \cite{Schuette}  and 
$f(0.8,0,12)\approx f(0.8,0,18)\approx 13$ \cite{Rybakov} for the decay  into the helical and conical state, respectively. Using $J\sim k_B  T_c$ and $N_s \sim 300$ for an order-of-magnitude estimate, we 
obtain $\frac{4 A^2}{D}\sim \frac{J N_s}{2 \pi} \sim 150\,$meV. Using this value, we conclude that the predicted activation energies are about an order of magnitude larger than the measured ones.

\clearpage

\end{document}